\documentclass[10pt,a4j]{article}
\usepackage[dvips]{graphicx}
\usepackage{mathrsfs}
\usepackage{amsmath,amsthm,amssymb}
\usepackage{wrapfig}
\usepackage[dvips]{color}
\usepackage{cite}
\usepackage{framed}
\definecolor{shadecolor}{gray}{0.80}
\DeclareMathVersion{chem}
\SetSymbolFont{letters}{chem}{OT1}{cmr}{m}{n}

\begin{document}

\renewcommand{\labelitemi}{}
\renewcommand{\thefootnote}{$\dagger$\arabic{footnote}}

\begin{flushright}
\textit{The Excluded Volume Problem}
\end{flushright}
\vspace{3mm}

\begin{center}
\setlength{\baselineskip}{25pt}{\LARGE\textbf{Minor Amendment of the Local Free Energy}}
\end{center}
\vspace{-5mm}
\begin{center}
\setlength{\baselineskip}{25pt}{\large\textbf{On Excluded Volume Problem}}
\end{center}
\vspace*{5mm}
\begin{center}
\large{Kazumi Suematsu} \vspace*{2mm}\\
\normalsize{\setlength{\baselineskip}{12pt} 
Institute of Mathematical Science\\
Ohkadai 2-31-9, Yokkaichi, Mie 512-1216, JAPAN\\
E-Mail: suematsu@m3.cty-net.ne.jp,  Tel/Fax: +81 (0) 593 26 8052}\\[8mm]
\end{center}


\section{Local Free Energy}
Amendment is necessary for the classic theory of the local free energy\cite{Flory} to be applied in a more rigorous manner to real polymer solutions. As is well known, the classic theory for the local free energy does not take into consideration correctly the interaction between segments on the same chain, so that the classic theory is constructed on the basis of the \textit{pseudo} self-avoiding chains whose behavior is virtually identical to that of the ideal chain. This feature has been frequently criticized so far by physicists as a fundamental deficiency of the Flory theory.

Let us consider the three dimensional lattice where a site may be occupied by a segment or a solvent molecule. Let a site be surrounded by $z$ neighboring sites, and let there be $\delta x$ segments from different molecules in the volume element $\delta V$. The number of arrangements of those segments is
\begin{equation}
\Omega_{\delta V}=\prod_{i=0}^{\delta x-1}(z_{i}-1)(1-f_{i})
\end{equation}
where $f_{i}$ is the probability that a given cell is occupied by the polymer segments when $i$ segments are already put in $\delta V$, so that $f_{i}=i/\delta n_{0}$ where $\delta n_{0}=\delta x+\delta n_{1}$, $\delta n_{0}$ denotes the total number of sites and $\delta n_{1}$ the number of solvent molecules in $\delta V$. $z_{i}$ is a special number introduced in this amendment and is defined by $0\le z_{i}\le z$. The physical meaning of $z_{i}$ is as follows:

\vspace*{1mm}
\begin{minipage}[t]{0.95\linewidth}
\setlength{\baselineskip}{13pt}
There is finite probability that a given segment overlaps with the other segments on the same chain (the multioccupation problem). Such unphysical conformations must properly be removed by subtracting from the total number $\prod_{i}(z-1)$ of feasible conformations. This is possible, because the number of conformations is enumerable in principle. The subtraction can be achieved simply by reducing $z$ to $z_{i}$, so that $\prod_{i}(z_{i}-1)$ represents the total number of self-avoiding conformations. To date the numerical value of $z_{i}$ is unfortunately unknown, which however is not essential for the present purpose, as is verified below.
\end{minipage}

\vspace*{1.7mm}
\noindent By eq. (1) the local entropy becomes
\begin{equation}
\delta S=\delta S_{mixing}+\delta S_{melting}= k\, \log\, \Omega_{\delta V}=k \left\{\sum_{i=0}^{\delta x-1}\log\, (z_{i}-1)+\log\, \frac{\delta n_{0}!}{\delta n_{0}^{\delta x}(\delta n_{0}-\delta x)!}\right\}
\end{equation}
Applying the Stirling formula to the above equation, we have
\begin{equation}
\delta S_{mixing}+\delta S_{melting}\cong -k \left\{\delta n_{1} \log\,v_{1}+\delta x-\sum_{i=0}^{\delta x-1}\log\, (z_{i}-1)\right\}
\end{equation}
where $v_{1}=(\delta n_{0}-\delta x)/\delta n_{0}$ is the volume fraction of solvents. The melting entropy, $\delta S_{melting}$, can be obtained by putting $\delta n_{1}=0$ in eq. (3). Hence we have
\begin{equation}
\delta S_{mixing}=-k\, \delta n_{1} \log\,v_{1}
\end{equation}
which is exactly the Flory result. It turns out that all the self-avoiding terms are absorbed into the melting entropy, $\delta S_{melting}$. Noteworthy is the fact that the mixing entropy $\delta S_{mixing}$ of the \textit{pseudo} self-avoiding chains and solvent is exactly equal to that of the genuine self-avoiding chains and solvent; only the standard state must be altered as
\begin{equation}
\text{pure } pseudo \text{ self-avoiding chains}\hspace{5mm} \Rightarrow \hspace{5mm} \text{pure } genuine \text{ self-avoiding chains}\notag. 
\end{equation}
This unexpected result does not appear to have been fully recognized by theorists up to present. A close investigation however tells us that the result is by no means surprising, because by eq. (4) $\delta S_{mixing}$ is a function of the solvent fraction alone, but unrelated to the conformational properties of chains.

Adding the enthalpy term, $\delta H_{mixing}=kT\chi\delta n_{1}v_{2}$ to the above equation, we have the formula of the local free energy in the volume element $\delta V$
\begin{equation}
\delta F_{mixing}=\delta H_{mixing}-T\delta S_{mixing}=kT\left\{\log\,\left(1-v_{2}\right)+\chi v_{2}\right\}\delta n_{1}
\end{equation}
where $v_{2}=1-v_{1}$ is the volume fraction of the segments. Eq. (5) represents the free energy difference between the mixture of the self-avoiding chains and solvent, and the respective pure components. Eq. (5) has already been derived by Flory\cite{Flory}.

We realize that eq. (5) has deeper generality along with sound physical basis, and hence is applicable equally to the excluded volume problem over all concentration range.



\begin{thebibliography}{99}
\bibitem{Flory} 
P. J. Flory. Principles of Polymer Chemistry. Cornell University Press, Ithaca and London (1953).
\end{thebibliography}
\end{document}